# Composition of Cosmic Rays with the Energy more than $4\times10^{19}$eV

A.A. Mikhailov

Yu.G. Shafer Institute of Cosmophysical Research and
Aeronomy, 31 Lenin Ave., 677891 Yakutsk, Russia

## Abstract

Arrival directions of extensive air showers by using world data are considered. It is found that the shower distribution in zenith angle at the energies $E>10^{19}$eV and $E>4\times10^{19}$eV differs from each other. By our estimations, the shower with $E>10^{20}$eV at the Sugar array was not registered. The mass composition of very high-energy cosmic rays has been estimated. It is shown that $E>4\times10^{19}$eV cosmic rays are, most likely, super heavy nuclei with charge $Z>26$.

## 1. Introduction

The composition of cosmic rays is the important characteristic to solve a problem of their origin. To clarify this question, the muon shower component as the most sensitive to the change of primary cosmic ray composition can play the essential role. The analysis of the muon component of extensive air showers (EAS') by using AGASA array data (Japan) shows that in cosmic rays at $E>10^{19}$eV the light nuclei are dominated [1]. The results obtained at the Hires array (USA) by data of the shift rate of shower development maximum depending on the energy show that cosmic rays at $E\sim2.5\times10^{19}$eV consist of light nuclei, most likely [2]. The estimation of cosmic ray composition at the Yakutsk EAS array by the Cherenkov radiation points to the fact that cosmic rays at $E\sim3\times10^{19}$eV consist mainly of the protons also [3]. Unfortunately, in these papers to interpret experimental data the model calculations are used which consider NN – and $\pi$N – interactions of very high-energy particles whose cross-sections are extrapolated from the accelerator region. In this extrapolation the inaccuracies can be. The experiments are also difficult and errors are not excluded.

Here we suggest a new method to estimate the cosmic ray composition on the basis of clearly established experimental data.

## 2. Experiment

Fig.1 presents the distribution of showers with $E>10^{19}$eV in zenith angle $\theta$: a - Yakutsk, b – Haverah Park [4]. The number of showers is 458 and 144, respectively. The dashed line is the expected number of events in the case of the isotropic distribution of the primary radiation according to [5]. Pearson $\chi^2$ – criterion shows that between observed and expected numbers of showers there is the fairly good agreement. As seen in Fig.1, in the shower distribution with $E>10^{19}$eV the inclined showers are predominated, as it is expected in the case of the isotropy.

In fig.2 the shower distribution at $E>4\times10^{19}$eV is shown: a – Yakutsk, b – AGASA [6]. The number of showers is equal to 29 and 47. The dashed line is the expected number of events in the case of the isotropy. A comparison of the observed and expected distribution of the showers has been made also using $\chi^2$ -test. For Yakutsk array the observed number of showers does not contradict the expected number of showers (probability that χ$^2$ - value is exceeded is ~ 0.15). The same is observed on the data of array AGASA (fig. 2b), but the probability - ~ 0.07. If to unit these two distributions of showers, χ$^2$ - test shows, the probability that χ$^2$ - value is exceeded is ~ 0.03. Hence the observed number of showers contradict the expected number of events in a case isotropy. At that in an interval of angles 20° - 30° the observed number of showers exceeds the expected ones on 2.3σ (see also [7]), where σ - standard deviation from expected number of events.

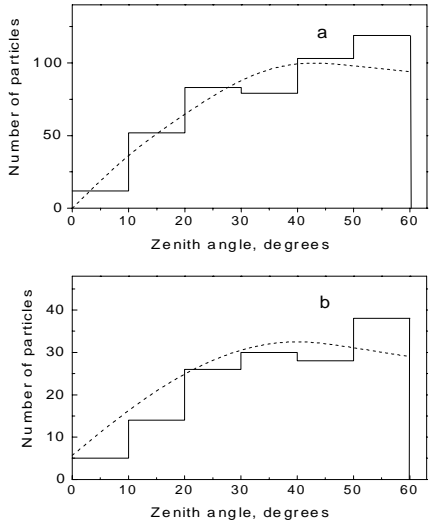

Fig.1. Distribution of showers with E>10$^{19}$ eV in zenith angle θ: **a**-Yakutsk, **b**-Haverah Park. The dashed line is the expected number of showers in the case of isotropy.

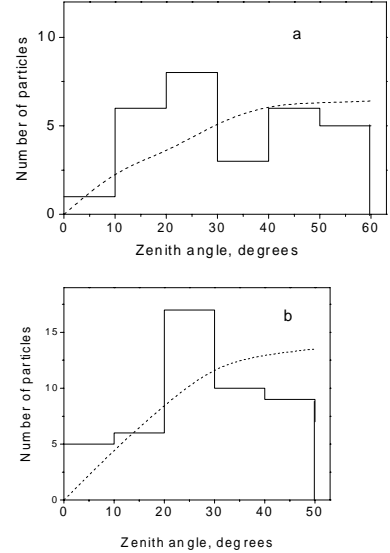

Fig.2. Distribution of showers with E>4×10$^{19}$ eV: **a**-Yakutsk, **b**-AGASA. The dashed line is the expected number of showers in the case of isotropy.

Thus, in the shower distribution at E>4×10$^{19}$eV the shower arrival maximum is in the range of average angles.

We consider the shower distribution in zenith angle by the SUGAR data. In [8] there are two variants to estimate of the shower energy: by the "Sydney" model and the "Hillas - E" model. Fig.3 shows the shower distributions by the "Sydney" model: the showers with E>10$^{19}$eV (a) and with E>4×10$^{19}$eV (b) but among them there are no showers with E>10$^{20}$ eV. These distributions are similar to the results obtained at the Yakutsk, Haverah Park and AGASA arrays (Fig.1,2) at corresponding energies. According to the "Hillas – E" model, the showers in Fig.3a have the energies higher than 4×10$^{19}$eV. In "Hillas – E" model the shower distribution (Fig.3a) is contradicted the data of the above arrays (Fig.2). On this basic, one can conclude that the estimation of the shower energy by the "Hillas – E" model is not correct, or according to "Sydney" model shower with E>10$^{20}$eV was not registered [8].

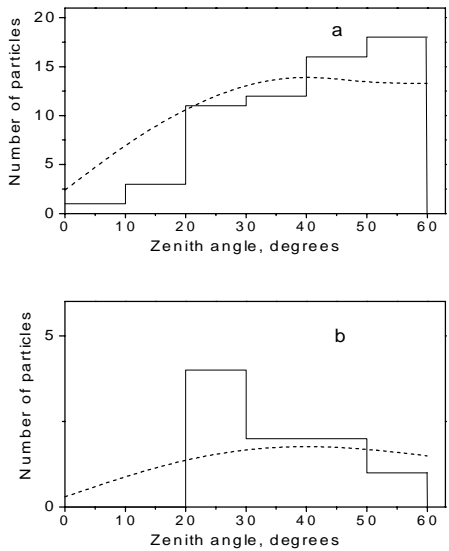

Fig.3. Distribution of showers by using Sugar array data. The "Sydney" model: **a**-E>10$^{19}$eV, **b**-E>4×10$^{19}$eV. The "Hillas-E" model: **a**-E>4×10$^{19}$eV. The dashed line is the expected number of shower in the case of isotropy.

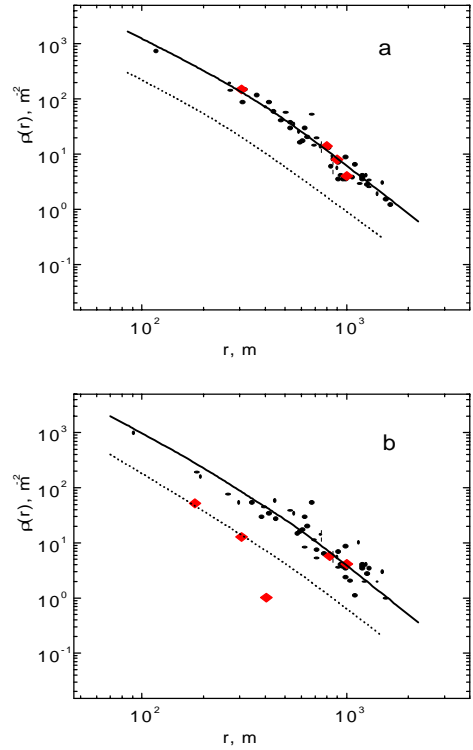

Fig.4. Particle density ρ(r) versus the distance r to a shower core: **a**-$E_1=1.2\times10^{20}$eV, **b**-$E_2=2\times10^{19}$eV, •-electrons and photons, ♦-muons, the solid and dashed lines are the expected densities of the electron-photon component and muons.

In order to clarify why the number of inclined showers at $E>4\times10^{19}$eV is small in comparison with the expected value (Fig.2), we analyze these showers.

Fig.4, demonstrates as an example of all data, the electron-proton and muon components of two inclined showers with angles and energies: $\theta_1=58.7°$, $E_1=1.2\times10^{20}$eV; $\theta_2=54.5°$ and $E_2=2\times10^{19}$eV. These showers are registered on May 7, 1989 and December 2, 1996 at the Yakutsk EAS array. The axes of the two showers are inside the array perimeter. As seen in Fig.4a, the particle densities in the scintillation detectors (registration threshold of electrons and photons is 3 MeV) and in the muon detectors (threshold is 1 GeV) become equal, i.e. the shower with $E_1=1.2\times10^{20}$eV consists of muons only. The shower with $E_2=2\times10^{19}$eV at the same zenith angle θ has the electron-photon and muon components (Fig.4b). The fact that a portion of muons in the inclined showers increases with the energy is established over all data in [9].

## 3. Discussion

Thus, two facts have been established at $E>4\times10^{19}$eV: 1) the shower arrival maximum is in the range of average angles, i.e. the relatively small number of inclined showers is observed; 2) the muon component in the inclined showers is beginning to predominate and at $E\sim10^{20}$eV it dominates as compared with other components. This facts can be interpreted as the change of the mass composition of galactic cosmic rays at $E>4\times10^{19}$eV to the side of more heavy nuclei.

The qualitative picture of the shower development is: a heavy nucleus interacts with air atoms in relatively high layer of the atmosphere in comparison with more light nuclei and disintegrates on the nucleons. The nucleons create the showers of small energy, and the inclined showers of relatively smaller energies are apparently absorbed stronger. Therefore a deficit of inclined showers takes place (Fig.2). However, the muon component begins apparently to dominate in the inclined showers of relatively high energies as compared with the electron-proton component. On this basis it may be concluded that the mass composition of cosmic rays with $E>4\times10^{19}$eV is more heavy than at $E\sim10^{19}$eV by a value:

$$A_1 > E_1/E_2 \times A_2 , \qquad (1)$$

where $A_1$ and $A_2$ is the atomic mass at $E_1$ and $E_2$ respectively.

Earlier we concluded that cosmic rays with $E>4\times10^{19}$eV were most likely galactic [7,10]. In [11,12] it is shown that cosmic rays at $E\sim10^{18}$ - $10^{19}$eV are most likely the iron nuclei. Thus, cosmic rays with $E>4\times10^{19}$eV are galactic and apparently more heavy than the iron nuclei.

Because of the mass composition change of the primary radiation, the lateral distribution function (LDF) of particles density in the showers can be varied. Hence the estimation of the shower energy is also changed (shower energy at the Yakutsk array is estimated by the particle density of electron-proton component at the distance of 600 m from a shower core). In [13] the overestimation of the shower energy $E_1=1.2\times10^{20}$eV was carried out in the assumption that the LDF of particles density of showers at very high energy varied and was obtained that this shower has $E_1=6\times10^{19}$eV. Substituting its value in (1), we obtain $A_1>168$, i.e. cosmic rays with $E>4\times10^{19}$eV are possible heavier than lanthanum nucleus (this is a preliminary result).

Thus, it is open question how many the composition of cosmic rays at $E>4\times10^{19}$eV more heavy than iron nucleus.

The clusters with $E>4\times10^{20}$eV [14] can be explained by fragmentation of super heavy nuclei in a result its collision with interstellar medium (another possible for form clusters is a spontaneity decay of nuclei). One of properties of such forms the clusters it will be the growth of the number of clusters with the energy. The reason is that the lifetime of nuclei will be increases with the energy and there is the probability that they decompose far from the their sources, i.e. nearer to the Earth. Experimental data points to the growth of the number of clusters with energy [15], i.e. this data confirm the above hypothesis.

## 5. Conclusion

In conclusion it may be said that very high-energy cosmic rays are galactic and consist of the nuclei heavier than an iron nucleus.

The Yakutsk EAS array was supported by the Ministry of Training of the Russian Federation, project no.01-30.